\documentclass[twocolumn,10pt]{article}
\usepackage[a4paper,margin=1in]{geometry}
\usepackage[final]{microtype}

\usepackage{packages}
\usepackage{commands}

\newcommand\imgext{%
    pdf%
}

\begin{document}

    
    \title{%
        Squeezed-vacuum bosonic codes
    }
    
    \author[1]{N. Gutman}
    \author[2]{E. Blumenthal}
    \author[2]{S. Hacohen-Gourgy}
    \author[1]{A. Orda}
    \author[1]{I. Kaminer}
    
    \affil[1]{%
        Viterbi Department of Electrical and Computer Engineering
    }
    \affil[2]{%
        Department of Physics
    }
    \affil[ ]{%
        Technion --- Israel Institute of Technology, Haifa 32000, Israel%
    }
    
    \date{November 2025}
    
    \maketitle
    
    \begin{strip}
        \vspace{-6em}
        \paragraph{Abstract:}
We introduce a family of \bosonic{} quantum error-correcting codes %
built as a rotation-symmetric superposition of squeezed vacuum states, %
which promise protection against both loss and dephasing noise channels.
The robustness of these ``squeezed-vacuum codes'' arises from being arranged at evenly spaced angles in phase-space, and simultaneously in evenly spaced photon-number support %
\({ n \equiv {2k} \! \pmod {2m} }\). %
We present simple preparation circuits
for general ``\(m\)-legged'' codewords using sequences of conditional rotations. %
The performance of these codes is evaluated against loss and dephasing noises using the Knill--Laflamme violation function and benchmarked against cat codes. %
As the number~\(m\) of squeezed-vacuum states in a code increases, the code exhibits improved loss tolerance at the cost of higher dephasing sensitivity. 
We outline implementations in circuit QED and trapped-ion platforms, where high-fidelity Gaussian operations and conditional controls are available or under active development. 
These results help establish squeezed-vacuum codes as practical, hardware-ready, members of the \bosonic{} codes class.

    \end{strip}



    \section{Introduction}
\label{sec:intro}

\vspace{-2em}

Quantum computers promise exponential speed increases for certain computational tasks, but their fragile quantum states are
extremely susceptible to noise.  Quantum error-correcting codes (QECCs)---beginning with
Shor's nine-qubit code~\cite{Shor1995} and formalized in the 
stabilizer framework~\cite{Gottesman1997}---protect logical quantum information %
by distributing it across a larger physical Hilbert space, enabling detection and correction of likely error
processes. Topological codes, such as the surface code~\cite{Kitaev1997,Fowler2012}, %
now dominate the road-maps of
large-scale qubit processors, owing to their relative simplicity, yet at the cost of cumbersome logical operations \cite{latticeSurgery} that require substantial qubit overhead.

An attractive alternative is to encode information in a \textit{\bosonic{}} mode, i.e.~the 
infinite-dimensional Hilbert space of a single quantum harmonic oscillator. 
Because multiple indistinguishable excitations can occupy one resonator, 
\bosonic{} codes can achieve the required redundancy by distributing information %
along its infinite-dimensional Hilbert space with surprising simplicity, %
while exploiting high-$Q$ cavities for long coherence~\cite{Milul2016PRB,WeizmannHighQ2023}. 
These \bosonic{} encodings naturally reside in the continuous-variable (CV) model of 
quantum computation, where quantum information is carried by the conjugate 
quadratures $(q, p)$ of an oscillator. 
\citet{LloydBraunstein1999} %
proved that Gaussian operations --- phase-space displacement $D(\alpha)$ and 
squeezing $S(r)$ --- must be supplemented by at least one non-Gaussian element to achieve %
universal CV quantum computation. 
\begin{figure}[H]
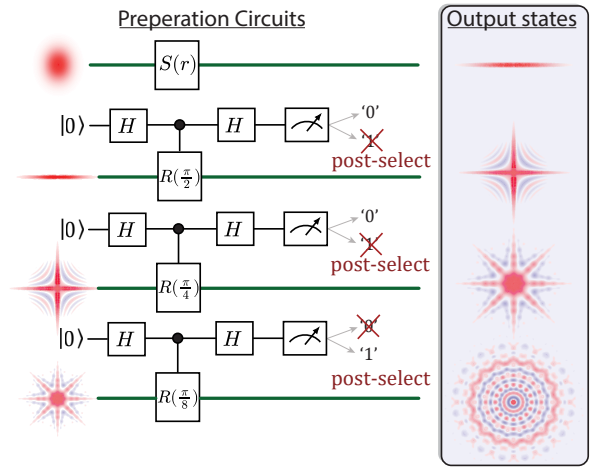

\centering
    \begin{overpic}[width=0.49\textwidth]{figures/fig00.\imgext}
    \end{overpic}
\vspace{-2em}
\caption{%
    \textbf{Generation of multi-legged squeezed states}: %
    We present a new family of quantum error-correcting \bosonic{} codes built from squeezed-vacuum states. %
    Each codeword can be created by a sequence of single-qubit gates, conditional-rotations and post-selection. %
    Each resulted \bosonic{} state can be fed back to generate a codeword with a larger superposition of squeezed-vacuum states. %
    Selecting different measurement results, generates different members of the family. In this example, the $8$-legged logical-$1$ codeword is generated.
}
\label{fig:preparation_concept}
\end{figure} %

Continuous-variable quantum error correcting-codes (CV-QECC)~\cite{chuangBosonicQuantumCodes1997} 
leverages \bosonic{} codes to protect against dominant noise channels, %
with landmark theoretical proposals including 
cat codes~\cite{Cochrane1999,Leghtas2013}, %
the Gottesman--Kitaev--Preskill (GKP) code~\cite{GKP2001}, 
and finite-energy binomial codes~\cite{Michael2016}. 
Advances over the past two decades have gradually established a unified description of CV-QECC \cite{ErrorCorrectionZoo,braunstein2005quantum,AlbertBosonicKrausNoise}. Such a general treatment enabled scientists to capture many useful encodings as special cases of the much larger class of \textit{rotation-symmetric bosonic codes}~\cite{Grimsmo2020}.

Experimental realizations of these codes have achieved significant milestones: 
Cat codes demonstrated break-even quantum error correction in superconducting cavities~\cite{Ofek2016}. %
Binomial codes showed universal gate operations and error correction in superconducting cavities~\cite{Hu2019}. %
GKP codes were first realized on trapped-ion platforms~\cite{Fluehmann2019} %
and demonstrated QEC with \(\sim \!\! 3\times\) longer logical coherences~\cite{deNeeve2022}; %
later implementations surpassed break-even %
and extended logical lifetimes to \(\sim \!\! 2\times\) that of their physical elements in superconducting circuits~\cite{Sivak2023};
and were recently created on-chip in integrated photonics~\cite{XanaduGKP2025}.

Despite impressive progress, a practical gap remains: %
existing \bosonic{} encodings rely on fundamentally different non-Gaussian resources %
(e.g., conditional displacement or Kerr nonlinearity), %
which are not uniformly available at high fidelity across platforms. %
Moreover, compact, rotation-symmetric encodings with finite energy that achieve 
useful number and phase distances without long gate sequences are scarce. 
This motivates us to seek CV codes that leverage simple Gaussian structure while minimizing reliance on 
demanding non-Gaussian resources.

Here we introduce a new \emph{family of bosonic error-correcting codes} %
built from superpositions of vacuum states squeezed along evenly-spaced axes. 
The resulting codewords are simultaneously rotation-symmetric and also belong to the \textit{number-phase} subclass~\cite{Grimsmo2020} %
with sparse Fock support. 
The simplest $2$-legged member achieves number code-distance~$2$ (i.e., loss events involving different numbers of photons map to disjoint subspaces) %
and an angular separation of $\pi/2$ between primitive squeezed states in phase-space, %
offering protection comparable to the $4$-legged cat code. %
We find that a controlled-squeezing operation --- %
squeezing the \bosonic{} mode conditioned on a 
qubit's state~\cite{blumenthalSingleTwomodeSqueezing,delgrossoControlledsqueezeGateSuperconducting2025,millicanEngineeringContinuousvariableEntanglement2025} %
--- can generate this $2$-legged state in a single step. 
On top of this, we develop preparation protocols for all members of the code-family, %
based on conditional rotations and single qubit logic.
We benchmark performance under photon loss and dephasing, comparing against cat codes.
We then provide details on controlled squeezing across platforms and discuss implementation considerations.

The remainder of the paper is organized as follows:
In \cref{sec:new-family} we introduce the squeezed-vacuum code family as a subclass of rotation-symmetric \bosonic{} codes and describe its structure and distance properties. Next, \Cref{sec:prep} presents probabilistic and deterministic circuits for code preparation.
Then, \Cref{sec:numerics} sets up loss and dephasing noise models and benchmarks the codes using a Knill--Laflamme cost, comparing against cat codes. 
Finally, \Cref{sec:discussion} addresses the experimental feasibility (circuit QED and trapped-ion platforms) and provides guidelines for choosing code parameters; additional derivations and analysis appear in the Supplementary Material (\cref{sec:supp:code-derivation,sec:supp:kl,sec:supp:ideal_phase_number}).

    
\section{A New Family of Codes}
\label{sec:new-family}

\begin{figure*}[htb]
\centering
\includegraphics[width=0.98\textwidth]
{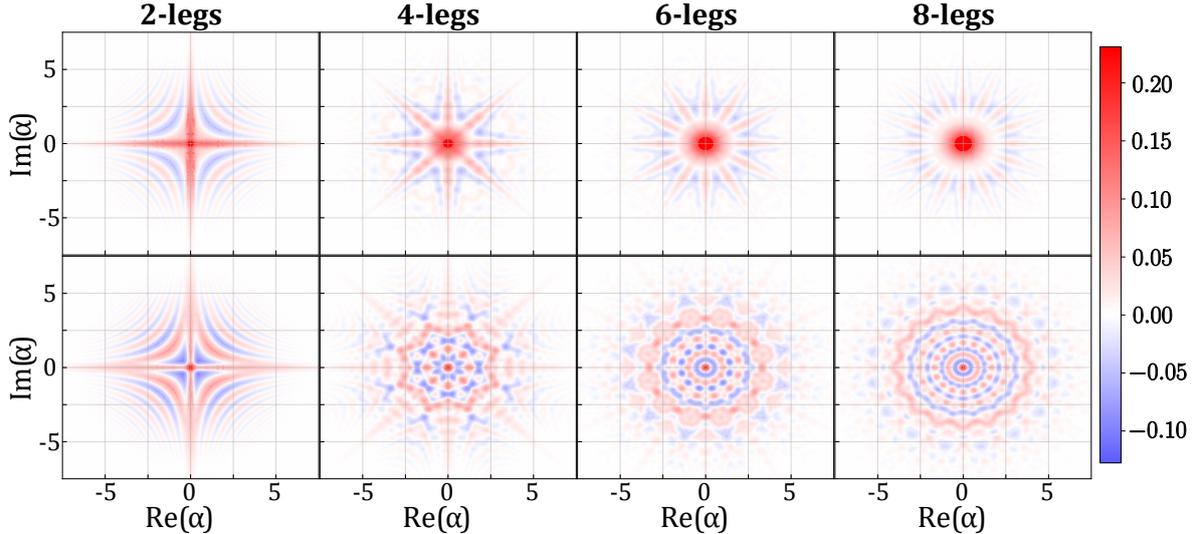}
\caption{%
    \textbf{The family of multi-legged squeezed \bosonic{} codes}: %
    Wigner functions %
    \cite{wignerFunction} %
    of $m$-legged squeezed codes for $m\!=\!2\ldots8$ at fixed squeezing strength $r=1.5$. 
    For each $m$, 
    we show $2$ logical codewords.
    Increasing the number of legs increases $m$-fold rotation symmetry and protection against particle loss. %
    The upper-left Wigner function was previously shown \cite{NICACIO20104385} but had not been analyzed there as a \bosonic{} code. 
}
\label{fig:wigner-family}
\end{figure*}

\subsection{Motivation: Superpositions of Gaussian States as Bosonic Codes}
Some of the simplest schemes that place Gaussian states in a superposition can protect against photon loss and dephasing~\cite{Grimsmo2020}.
A paradigmatic example is \emph{Schrödinger cat code}, which stores a qubit in an $m$-legged superposition of coherent states
\cite{Cochrane1999,Leghtas2013}.
The legs are arranged equidistantly on a circle in phase space, endowing the state with an $m$-fold rotational symmetry that translates into resilience to particle-loss channels.
Thus, by only manipulating Gaussian resources (displacements, squeezers, and phase shifters), one can control a rich highly-nonclassical code space. Our code family is such a surprisingly simple set.

\textbf{Cat codes as a family.}
Rather than a single encoding scheme, cat states form a continuum parameterized by their mean photon number $|\alpha|^2$ and an integer $m$ specifying the number of coherent-state "legs".
\cref{fig:two-leg-circuit}\figCap{a2} depicts the familiar two-legged cat. Increasing $m$ reduces the angular separation between the legs, thus improving particle-loss tolerance at the cost of less resilience to dephasing. This increased susceptibility to dephasing noise can be intuitively attributed to the larger number of states taking the same volume in phase-space. 

\subsection{Required operations}
The canonical cat-code preparation uses a \emph{conditional displacement}, in which an ancilla qubit coherently displaces the oscillator to amplitudes $\pm\alpha$. We now build on this idea by replacing displacement with \emph{conditional squeezing} \cite{blumenthalSingleTwomodeSqueezing, delgrossoControlledsqueezeGateSuperconducting2025,millicanEngineeringContinuousvariableEntanglement2025}.

Recall the single-mode squeezing Hamiltonian~\cite{wallsSqueezedStatesLight1983}
\begin{equation}
H_{\mathrm S}(\kappa)
= \frac{i}{2}\bigl(\kappa\,a^{\dagger 2}-\kappa^{*}\,a^{2}\bigr),
\label{eq:sqz-ham-general}
\end{equation}
whose conjugate coefficients ensure anti-Hermiticity of the generator. This Hamiltonian generates (for interaction duration \(t\)) a squeeze of magnitude \(
    {
    r=|\kappa| t
    }
\),
corresponding to the unitary operator,
\begin{equation}
S(\zeta)
= \exp\!\Bigl[\tfrac12\bigl(\zeta^{*} a^{2}-\zeta\, a^{\dagger 2}\bigr)\Bigr],
\label{eq:sqz-unitary-phi}
\end{equation}
with \(\zeta=r e^{i\phi}\).
Here, we would like to follow the notion of \emph{primitive-states} in phase-space given in \cite{Grimsmo2020}. Our primitive state is a single squeezed vacuum state~\cite{wallsSqueezedStatesLight1983}, elongated along direction $\theta$.
we relate \(\theta\) to \(\phi\) with
\begin{equation}
\phi(\theta)=2\theta+\pi \ \ (\mathrm{mod}\ 2\pi),
\label{eq:theta-to-phi}
\end{equation}
which yields, e.g., \(\theta=0 \text{ or }\theta=\pi \;
\Rightarrow
\;
\phi=\pi\).
We then define the direction-labeled squeezing unitary as
\begin{equation}
S(r,\theta)\ \equiv\ S\!\bigl(r e^{i\phi(\theta)}\bigr) \,.
\label{eq:S_r_theta}
\end{equation}
%
Finally, we define the conditional squeezing unitary as
\begin{equation}
\begin{split}
    CS(r;\theta_0,\theta_1)
    &= 
    \ket{0}\!\bra{0}\otimes S(r,\theta_0)
    \\&+
    \ket{1}\!\bra{1}\otimes S(r,\theta_1),
\end{split}
\label{eq:conditional-squeezing}
\end{equation}
which applies squeezing that elongates vacuum along \(\theta_0\) (if the ancilla is \(\ket{0}\)) or along \(\theta_1\) (if it is \(\ket{1}\)). 
Practical implementations \cite{blumenthalSingleTwomodeSqueezing,delgrossoControlledsqueezeGateSuperconducting2025,millicanEngineeringContinuousvariableEntanglement2025} %
of such an operation across multiple platforms (circuit QED, trapped ions) are discussed in \cref{sec:discussion}.

\subsection{Squeezed-vacuum Codes}
Here we present a new family of \emph{Squeezed-vacuum} codes defined by a coherent superposition of $m$ %
differently oriented squeezing operations on the vacuum state.
For an even integer $m>0$ and rotation-symmetry index $k \in[0, m\!-\!1]$ we set
\begin{equation}
\label{eq:psi_k}
    |\psi_k\rangle
    \propto 
    \sum_{j=0}^{m-1} 
    e^{
        i \frac{2\pi j}{m} k
    } \,
    S(r, \frac{\pi j}{m}) \,
    \ket{\vac}.
\end{equation}
In Fock space, this state occupies photon numbers \({
    n\equiv 2k \! \pmod{ 2m}
}\), yielding natural logical codewords
$|0_L\rangle := |\psi_0\rangle$ and $|1_L\rangle := |\psi_{m/2}\rangle$ %
whose number distributions are interleaved by $\Delta n = m$ %
(%
    See \cref{sec:supp:code-derivation} for a derivation
)
. %
The code distance is therefore $d = m$ against single-photon loss events.

This sets the stage for the definition of a new family of single-mode \bosonic{} error correction codes, as follows.
For two parameters $m$ and $r$ --- the even integer "number of legs" and the squeezing strength, correspondingly --- two logical states are constructed from an equal (sometimes phased) superposition of squeezed states according to \cref{eq:family_def}:
\begin{equation}
\begin{split}
    |0_L\rangle
    & \propto 
    \sum_{j=0}^{m-1} 
    S(%
        r, 
        \frac{\pi j}{m}
    ) \,
    \ket{\vac},
    \\
    |1_L\rangle
    & \propto 
    \sum_{j=0}^{m-1} 
    (-1)^{j} 
    \,
    S(%
        r, 
        \frac{\pi j}{m}
    ) \,
    \ket{\vac},
\end{split}
\label{eq:family_def}
\end{equation}
with a trivial normalization factor that keeps each state normalized to $1$.
Since single-primitive squeezed state are already limited to even Fock numbers only. %
codes with odd number of legs $m$ are generally possible, but result in codewords without sparse Fock-number occupancy.
A closed-form expression for the Fock-basis representation is derived in the Supplementary Material (\cref{sec:supp:code-derivation}).

This family of codes --- as a sub-family of rotation-symmetric bosonic codes --- follows the same intuition from \cite{Grimsmo2020} that more primitives (more legs / larger $m$) results in a code that is more robust to loss but more susceptible to dephasing. We explore this notion numerically in \cref{sec:numerics}.
Additionally, this code is an "approximate number-phase code"~
\cite{Grimsmo2020} %
and we approach the "ideal number-phase code" at the limit of \(
    r \rightarrow \infty
\) (See discussion in \cref{sec:supp:ideal_phase_number}).

Universal quantum computation using our codes would require multiple \bosonic{} modes and the ability to generate entangling quantum gates between them. To that end, one can use a controllable cross-Kerr interaction %
\(
    a^{\dagger}a \otimes    a^{\dagger}a
\) %
that enables the generation of the (entangling) $CZ$ gate ($diag(1,1,1,-1)$) \cite{Grimsmo2020}. This operation becomes viable across multiple platforms \cite{crossKerr_cQED_Hu,crossKerr_cQED_Liu,crossKerr_ions}.

While working on this paper, we discovered \citet{korolevErrorCorrectionUsing2024}, which beautifully discusses squeezed Fock states as a QECC source, which for $n{=}0$ (Vacuum state) is the special case of $m{=}1$ instance of our family%
\footnote{%
    or $m{=}2$ in the Hadamard basis
}; %
and %
\cite{hopePreparationConditionallysqueezedStates2025} %
which studies the preparation and error correcting properties of the %
$m{=}2$ codewords in the formalism introduced in this work.

    \section{Preparation Circuits}
\label{sec:prep}

\begin{figure*}[htb]
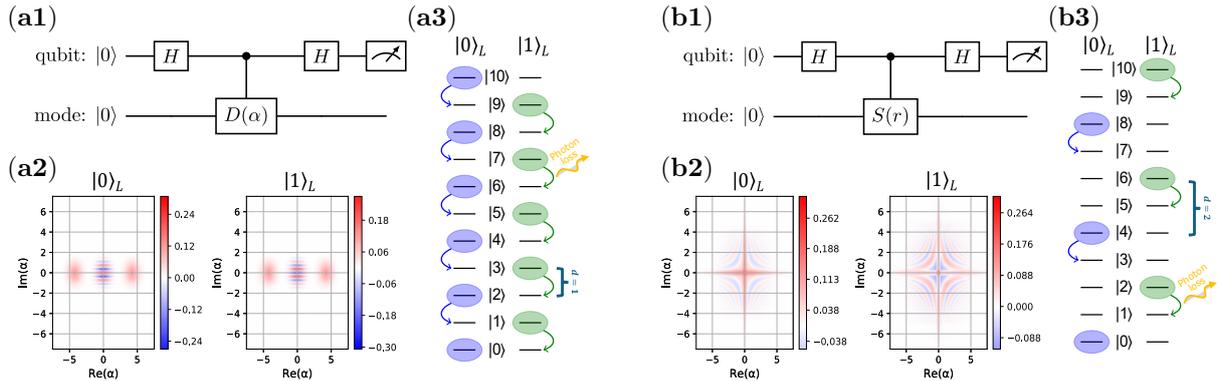

\centering
    \begin{overpic}[width=1\textwidth]{figures/fig02.\imgext}
        \put( 0, 30.5){\figCap{a1}}
        \put( 0, 18.0){\figCap{a2}}
        \put(33, 30.5){\figCap{a3}}
        \put(54, 30.5){\figCap{b1}}
        \put(54, 18.0){\figCap{b2}}
        \put(86, 30.5){\figCap{b3}}
    \end{overpic}
\vspace{-2em}
\caption{%
    \textbf{Generating scheme of multi-legged squeezed codes vs cat codes}: %
    Comparing a conventional circuit for generating the %
    $2$-legged Schrödinger's cat state using a conditional-displacement gate $CD(\alpha)$ \figCap{a1} %
    and our $2$-legged squeezed code using a conditional-squeezing gate $CS(r;0,\frac{\pi}{2})$ \cref{eq:conditional-squeezing} \figCap{b1}. Wigner functions %
    \cite{wignerFunction} %
    of the resultant states are shown in \figCap{a2,b2}.     
    The support of the codes in Fock space is also compared \figCap{a3,b3}: %
    While the 2-legged cat code has a number distance of $d{=}1$ \figCap{a3} with respect to photon loss events, %
    our code has a number distance of $d{=}2$, %
    meaning that single-photon loss events map to disjoint subspaces \figCap{b3}. %
    The number distance increases linearly for both codes as the number of legs $m$ increases.
    Circuits drawn with \cite{quantikz2018tutorial}.
}
\label{fig:two-leg-circuit}
\end{figure*}

In this section we discuss probabilistic and deterministic methods for the generation of the \bosonic~ codewords in the squeezed-vacuum family. %
We start with the simplest circuit, namely generating the code words for the $2$-legged code, in a probabilistic manner. We then generalize to arbitrary even $m$-legged variants of the code. We also discuss the necessary operators needed to make these processes deterministic.

\subsection{\titleMath{2}-legged codewords}
\label{sec:Probabilistic_Preparation}

A minimal circuit to probabilistically prepare the $m{=}2$ codeword pair uses a Hadamard--Controlled-Squeezing--Hadamard (\(
    H{-}CS{-}H
\)) %
sequence on a single ancilla-qubit coupled to a bosonic mode initially in the vacuum state; the probabilistic nature of this method arises from the final $Z$-basis measurement, collapsing the mode state to either an even or an odd superposition of squeezed states. \cref{fig:two-leg-circuit} illustrates that protocol and compares it to a similar sequence that generates the $2$-legged cat code. 

Although it might seem that the probability of getting either of the logical code states is equally distributed $50{-}50$, this is \textbf{not} the case.
Consider the case of $r{=}0$, equivalent to not performing squeezing at all. The logical-$0$ state reduces to the vacuum state, but the logical-$1$ state cannot be achieved. Therefore, we should expect a probability that scales somehow with the squeezing strength. 
In fact, the probability of getting the logical state $L$ (\(
    L \in \{0, 1\}
\)), given the squeezing strength $r$, is:
\begin{equation}
    \operatorname{prob}(L \, | \, r)
    =
    \frac{1}{2} 
    + 
    \frac{
        (-1)^L
    }{
        2 \cosh{r}
        \sqrt{
            \tanh^{2}{r} + 1
        } 
    } \;.
\label{eq:probability_of_preparation}
\end{equation}
As $r$ increases, the probabilities %
approach a uniform distribution. %

In most cases, such a probabilistic nature can be tolerated as long as we add post-selection on the measurement result.

\subsection{General \titleMath{m}-legged codewords}
\label{sec:Deterministic_Preparation}

Extending to $m{>}2$ by simply repeating the process in \cref{sec:Probabilistic_Preparation}
would not work. Simply put, squeezing operations do not stack on top of each other: instead of creating more superpositions of primitive squeezed-vacuum states, more conditional-squeezings on top of an already squeezed states would only rotate and stretch the existing states.

As shown in %
\citet{blumenthalSingleTwomodeSqueezing, delgrossoControlledsqueezeGateSuperconducting2025}, %
controlled-squeezing promises universal control over the joint system of %
qubit and harmonic oscillator mode, once it is combined with displacement and %
squeezing on the mode and with full single-qubit control. %
This means that we can generate any unitary acting on the joint Hilbert %
space of the qubit and the quantum harmonic oscillator, %
with accuracy that increases with the number of gates.

Two such unitaries that we can generate are: 
the conditional-rotation, %
\(
    CR(\theta)
\) %
which rotates the mode in phase space by $\theta$ if the state of the qubit (possibly in a superposition) is $\ket{1}$ (\cref{eq:conditional-rotation}); %
and the logical-$X$ gate
\(
    X_L
\)%
, which acts as the Pauli-$X$ gate in the $m$-legged code-space, %
transforming the logical-$0$ state into a logical-$1$ state, and vice-versa (\cref{eq:logical-x}).

\(
    CR(\theta)
\) %
can be written as %
\begin{equation}
    CR(\theta)
    = 
    \ket{0}\!\bra{0}\otimes \id
    +
    \ket{1}\!\bra{1}\otimes R(\theta),
\label{eq:conditional-rotation}
\end{equation}
where \(
    R(\theta) 
    =
    e^{
        i \theta a^{\dagger} a
    }
\) rotates the state in phase-space by $\theta$ counter-clockwise. %
This operation can also be achieved easily in contemporary systems %
with dispersive coupling between a qubit and an oscillator, %
such as circuit QED and trapped ions. %
The dispersive interaction \(
    H_\mathrm{Disp}=\frac{\chi}{2}\sigma_z a^\dagger a
\) %
generates conditional rotations of $\theta=\chi T$ during an idling time of $T$~\cite{SNAPGate2015}.

\(
    X_L
\) %
can be written as %
\begin{equation}
    X_L
    = 
    \ket{0_L}\!\bra{1_L}
    +
    \ket{1_L}\!\bra{0_L}
    +
    \id_{\perp} \,,
\label{eq:logical-x}
\end{equation}
where \(
    \id_{\perp}
\) is the identity operator on the orthogonal complement of the code space. %
Note that as the squeezing strength $r$ increases, %
the logical pauli-$X$ operation can be approximated by the operator \(
    \sum_{n=0}^{\infty} \ket{n}\!\bra{n+m}
\), 
since the logical codewords approach ``\emph{ideal number-phase codes}''%
\cite{Grimsmo2020}. %
See related discussion in~\cref{sec:supp:ideal_phase_number}.

With these gates, we can establish the following algorithm for the generation of \(m\)-legged code states, for \(m{=}2^k\) with a positive-integer $k$.
Starting with the qubit in $\ket{0}$ and the mode in $\ket{\vac}$, we first squeeze our mode by $r$, creating the squeezed primitive state $S(r,0)\ket{\vac}$. Then, we repeat the following sequence for \(\log_2(m)\) iterations:

Apply \( 
    (H \otimes \id)
    CR(\theta_j)
    (H \otimes \id)
\). %
At the $j^{\text{th}}$ iteration, 
\(
    \theta_j = 
    \frac{\pi}{2^j}
\). %
Then, measure the qubit. If this is the last iteration, the result of the measurement dictates whether we are in the logical-$0$ or logical-$1$ state.
Otherwise, only the $'0'$ result should be kept for the next iterations. One can either use post-selection (probabilistic) or a feed-forward conditional $X_L$ correction (deterministic) to keep the desired result. This algorithm is summarized in \cref{alg:det-prep-CR} and visually exemplified in \cref{fig:preparation_concept}.

\def\mode{\ket{\text{mode}}}
\def\CRphase{%
    CR_{\text{phased}}%
}

\begin{algorithm}[htbp]
\caption{Preparation of $2^k$-legged codewords using conditional-rotations}
\label{alg:det-prep-CR}
\begin{algorithmic}[1]
    \State \textbf{Inputs:} $r, \;k, \; \hat L$
    \Comment{$\hat L\in \{0,1\}$ is the desired codeword}
    \State Initialize bosonic mode in $\ket{\vac}$
    \State $\mode \leftarrow S(r,0)\ket{\vac}$
    \Comment{Apply squeezing}
    \For{$j = 1$ to $k$}
        \State $\theta_j \leftarrow \frac{\pi}{2^j}$ 
        \State $\ket{\psi} = \ket{0} \otimes \mode$ 
        \Comment{Initialize qubit in $\ket{0}$}
        \State \(
            \ket{\psi} \leftarrow 
            (H \! \otimes \! \id) \,
            CR(\theta_j) \,
            (H \! \otimes \! \id)
            \ket{\psi}
        \)
        \State Measure the qubit in the Z basis
        \State $L \leftarrow$ measurement result
        \Comment{ $L \in \{0,1\}$}
        \State $\mode \leftarrow \ket{\psi}$ traceover qubit
        \Statex \Comment{ deterministic or probabilistic part (see \hyperref[alg:note]{Note}):}%
        \If{$j < k$}  
            \State continue only if $L=0$ 
        \Else
            \State \textbf{return} $\mode$ if $L=\hat{L}$
        \EndIf
    \EndFor
\end{algorithmic}

\vspace{0.5em}
\begin{minipage}{0.95\linewidth}
\small
\textbf{Note \label{alg:note}}:  
If an undesired outcome $L$ occurs, two approaches are possible.  
In a deterministic protocol, one applies a corrective logical-$X$ operation (\cref{eq:logical-x}) on $\mode$.  
In a probabilistic protocol, one instead uses post-selection, i.e., discards the current run and repeats until the desired outcome is obtained.  

\end{minipage}
\end{algorithm}

This rather simple protocol can generate only codewords of the $m{=}2^k$-legged variants. To generate a general even $m$-legged state, we apply a single iteration of the previous protocol on an already $\frac m2$-legged state in an equal superposition \(
    \ket{e_{\frac m2}}
    =
    \sum_{j=0}^{\frac m2-1} 
    S(%
        r, 
        \frac{2\pi j}{m}
    ) \,
    |0\rangle
\). If $\frac m2 = 2^k$ for some $k$, then we can use the protocol from \cref{alg:det-prep-CR}, accepting only the logical-$0$ codeword (which is of an equal superposition as required); Otherwise, $\ket{e_{\frac m2}}$ can be achieved with another protocol, summarized in \cref{alg:det-prep-CR:general-m}.
For this protocol, we will introduce an operation that conditionally applies a rotation with phase, which we will write as 
\(
    \CRphase(\theta, \phi)
\) %
(\cref{eq:cr_phase}).
\begin{equation}
    \CRphase(\theta, \phi)
    = 
    \ket{0}\!\bra{0}\otimes \id
    +
    e^{i\phi} \,
    \ket{1}\!\bra{1}\otimes R(\theta).
\label{eq:cr_phase}
\end{equation}
This operation can be decomposed as a conditional rotation preceded by a phase gate on the qubit:
\begin{equation}
    \CRphase(\theta, \phi)
    = 
    \left(
        e^{i\phi\ket{1}\!\bra{1}} \otimes \id
    \right)
    CR(\theta),
\label{eq:cr_phase_decomposition}
\end{equation}
where the phase gate %
\(
    e^{i\phi\ket{1}\!\bra{1}}
    =
    \ket{0}\!\bra{0} + e^{i\phi}\ket{1}\!\bra{1}
\) %
acts only on the qubit.

\begin{algorithm}[htbp]
\caption{Preparation of $m$-legged squeezed states in equal superposition using conditional rotations}
\label{alg:det-prep-CR:general-m}
\begin{algorithmic}[1]
    \State \textbf{Inputs:} $r, \;m$
    \State Initialize bosonic mode in $\ket{\vac}$
    \State $\mode \leftarrow S(r,0)\ket{\vac}$
    \Comment{Apply squeezing}
    \State $\theta \leftarrow \frac{2\pi}{m}$ 
    \For{$j = 1$ to $m-1$}
        \State $\phi_j \leftarrow 
        \pi - \frac{
            2 \pi j
        }{
            m
        }$ 
        \State $\ket{\psi} = \ket{0} \otimes \mode$ 
        \Comment{%
            Initialize qubit in $\ket{0}$%
        }
        \State \(
            \ket{\psi} \leftarrow 
            (H \! \otimes \! \id) \,
            \CRphase(\theta, \phi_j) \,
            (H \! \otimes \! \id)
            \ket{\psi}
        \)
        \State Measure the qubit in the Z basis
        \State $L \leftarrow$ measurement result
        \Comment{$L \in \{0,1\}$}
        \State $\mode \leftarrow \ket{\psi}$ traceover qubit
        \Statex \Comment{%
            deterministic or probabilistic part (see \hyperref[alg:note]{Note}):%
        }%
        \State continue only if $L=0$ 
    \EndFor
    \State \textbf{return} $\mode$
\end{algorithmic}
\end{algorithm}

Python implementations of these protocols can be found in our code \cite{gitCode}.
    \section{Numerical Analysis}
\label{sec:numerics}


Errors in quantum systems are modeled by completely positive and trace-preserving (CPTP) linear maps (quantum channels). One convenient description is a channel's Kraus (operator-sum) representation:
\begin{equation}
    \mathcal{N}_{\gamma}
    (\rho)
    =
    \sum_{j} 
    K_j\,\rho\,
    K_j^\dagger
    ,
    \quad
    \sum_{j}
    K_j^\dagger 
    K_j = \id.
\end{equation}
This representation is not unique.

In applications related to quantum computation with bosonic codes, the dominant noise processes are loss%
\footnote{%
    colloquially called 'photon-loss,' even though the system may not be photon-based.%
} %
and dephasing, each with its infinite series of Kraus operators, given here as a function of a noise-strength $\gamma$~\cite{leviantQuantumCapacityCodes2022}:
\begin{equation}   
\label{eq:noise_channels:loss_dephasing:krauss_rep}
\begin{split}   
K_{j}^{{\text{loss}}}(\gamma)
=
\sqrt{
    \frac{
        \gamma^j
    }{
        j!
    }
}\,
(1-\gamma)^{
    \frac{n}{2}
}
\,
 a^j,
\\
K_{j}^{{\text{dephasing}}}(\gamma)
=
\sqrt{
    \frac{
        \gamma^j
    }{
        j!
    }
}\,
e^{
    -
    \frac{\gamma}{2}
    n ^2
}
\,
 n^j,
\end{split}
\end{equation}
where \(\gamma\) is the unitless noise-strength coefficients, \( a\) is the annihilation operator and \( n= a^\dagger a\). 
These two error channels commute~\cite{channelsCommute,leviantQuantumCapacityCodes2022}, thus allowing independent analysis.

The performance of a \bosonic~code with two orthonormal logical codewords %
\(|0_L\rangle\) %
and %
\(|1_L\rangle \) %
can be measured according to its Knill--Laflamme (KL) violation. The original KL conditions for exact correctability are
\(
    \langle i_L|
    K_a^\dagger
    K_b|j_L\rangle
    \propto
    \delta_{ij}
\)~\cite{KnillLaflammw}.
Exact satisfaction of these conditions yields a perfect code with 
\begin{equation}
\label{eq:numeric:perfect_codes_kl}
\begin{split}
    \langle 0_L| K_a^\dagger K_b|0_L\rangle
    &=
    \langle 1_L| K_a^\dagger K_b|1_L\rangle
    =1
    \quad
    \forall a,b \,,
\\
    \langle 0_L| K_a^\dagger K_b|1_L\rangle
    &=
    \langle 1_L| K_a^\dagger K_b|0_L\rangle
    =0
    \quad
    \forall a,b \,.
\end{split}
\end{equation}
Approximately satisfying them gives rise to realistic, experimentally accessible codes that correct errors only up to a given tolerance. %

One way to measure the performance of an approximate code is to sum the deviation from \cref{eq:numeric:perfect_codes_kl} for specific ket $\bra{i}$ and $\ket{j}$, along all Kraus operators in a given noise-channel. We call this the KL-violation of $\bra{i}$ and $\ket{j}$ over noise-channel $\mathcal{N}$. An exact formula is given in \cref{eq:KL_violation_formula}:

\begin{equation}
\label{eq:KL_violation_formula}
    V_\text{KL}(%
        \mathcal{N}%
     )_{i,j}
     =
     \sum_{a,b}
     \left|
         \delta_{ij}
         +
         (-1)^{\delta_{ij}}
         \bra{i}
         K_a^{\dagger} 
         K_b
         \ket{j} 
    \right|
     \,.
\end{equation}
Summing over all Kraus operators in an infinite series (\cref{eq:noise_channels:loss_dephasing:krauss_rep}) is impossible in numerical analysis, so a cut-off must be chosen. Choosing a cut-off in a way that keeps the results accurate up to a predetermined tolerance is discussed in \cref{sec:supp:kl}.

It is also possible to sum \(
    V_\text{KL}(%
        \mathcal{N}%
     )_{i,j}
\) %
over all logical $i,j$ states combinations --- into a single "KL cost function"~\cite{Reinhold2019}. This allows computerized optimization protocols to choose a specific code scheme for a given system under constraints and other considerations.
Even more methods to compare the performance of single-mode bosonic codes are emerging. For example, methods that take the recovery procedure into account~\cite{NearOptimalPerformance,AlbertBosonicKrausNoise}. There is a plethora of other options that we will not discuss in this paper.

\begin{figure}[htbp]
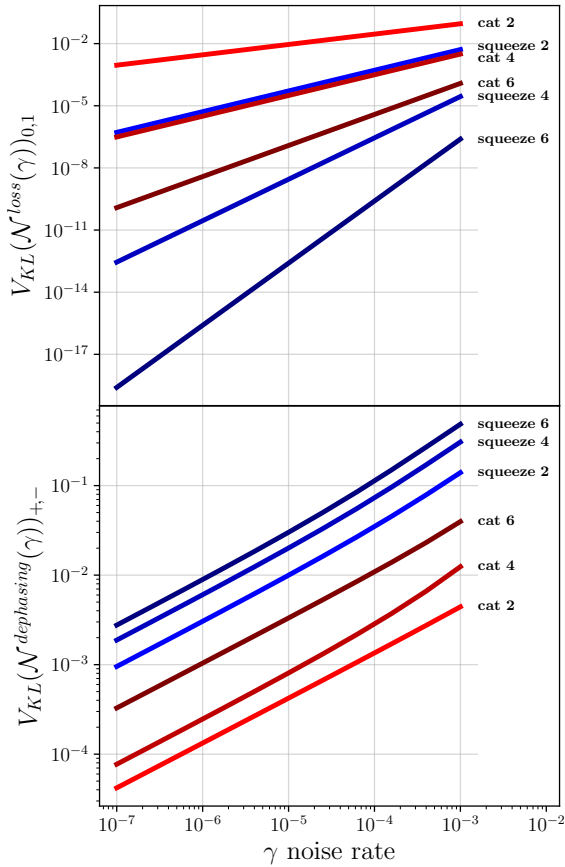

\centering
    \begin{overpic}[width=0.49\textwidth]{figures/fig_numerics1.\imgext}
    \end{overpic}
\caption{%
    \textbf{Analysis of error correction properties, comparing the squeezed- and cat- bosonic codes}: Plotting \(
        V_\text{KL}(%
            \mathcal{N}%
         )_{0,1}
    \) (\cref{eq:KL_violation_formula}) %
    for loss noise and \(
        V_\text{KL}(%
            \mathcal{N}%
         )_{+,-}
    \) %
    for dephasing noise, over strengths \(\gamma\).
    Exact KL conditions satisfactions occurs at \(
        V_\text{KL}(%
            \mathcal{N}%
         )
         = 0 
    \) hence smaller values indicate better code performance. 
    We compare our squeezed-vacuum codes to the established cat codes. %
    Here the squeezing strength and the displacement magnitude are chosen %
    so that both codes are compared with the same mean photons number %
    $\bar n=2$ %
    for $\ket{0_L}$
    (See \cref{eq:nk-mean} in~\cref{sec:supp:code-derivation}). 
    One can observe that increasing the number of legs protects against loss better at the cost of worse protection against dephasing.
}
\label{fig:numerical_results1}
\end{figure}


\cref{fig:numerical_results1} %
shows \(
    V_\text{KL}(%
        \mathcal{N}%
     )_{0,1}
\) %
for loss noise and \(
    V_\text{KL}(%
        \mathcal{N}%
     )_{+,-}
\) for dephasing noise, each over different noise-strengths \(\gamma\). %
The dual codewords \(
    \left\{
        \ket{+_L} = \frac{1}{\sqrt{2}}\left(
            \ket{0_L} + \ket{1_L}
        \right)
    ,
        \ket{-_L} = \frac{1}{\sqrt{2}}\left(
            \ket{0_L} - \ket{1_L}
        \right)
    \right\}
\) %
are more prone to dephasing error, so it is more fair to test code performance given those states.

Our numerical analysis can be replicated through our (freely-accessible) code \cite{gitCode}.

    \section{Discussion and outlook}
\label{sec:discussion}

\textbf{Platform readiness (circuit QED):}
Practical deployment of the squeezed-vacuum code depends on the availability of high-fidelity Gaussian operations (displacements, rotations, squeezing) together with conditional control. Circuit QED platforms are particularly promising: universal oscillator control via a weakly-dispersive ancilla has been established and scaled to fast, high-fidelity gates~\cite{Krastanov2015,Eickbusch2022}. In this toolbox, all ingredients for state preparation, manipulation, and readout are available—number-selective conditional rotations (SNAP-type gates)~\cite{SNAPGate2015}, high-quality single-mode squeezing~\cite{Eickbusch2022}, and, recently, proposals for a conditional-squeezing gate compatible with standard transmon–cavity hardware~\cite{blumenthalSingleTwomodeSqueezing}. %
These capabilities align naturally with our suggested preparation circuits.

\textbf{Platform readiness (trapped ions):}
Trapped-ion systems have demonstrated comparable levels of squeezing and even deterministic preparation of GKP logical states in the motional modes of a single ion~\cite{Fluehmann2019,matsos2024}. 
With access to sideband couplings (spin-dependent forces)~\cite{Haljan2005}, and optimal-control pulse shaping for spin–motion dynamics~\cite{matsos2024}, trapped-ion platforms can realize conditional Gaussian primitives—including state-dependent displacement~\cite{Haljan2005} and controlled-squeezing~\cite{Drechsler2020}—suggesting a viable route to implement the suggested code family.

\textbf{Creation of controlled squeezing:}
Recent proposals and demonstrations across platforms indicate growing readiness for the implementation of a controlled-squeezing operation~(\cref{eq:conditional-squeezing}).
In circuit QED, Blumenthal et al.~\cite{blumenthalSingleTwomodeSqueezing} propose using a Rabi-driven %
qubit dispersively coupled to one or two high-$Q$ oscillators to implement conditional single- and two-mode squeezing %
entirely within the cavity, predicting up to $\sim4$\,dB of single-mode squeezing in superposition and %
observing qubit-oscillator entanglement without additional nonlinear elements. %
Independently, Del Grosso et al.~\cite{delgrossoControlledsqueezeGateSuperconducting2025} introduce a %
controlled-squeeze gate in a resonator terminated by a SQUID (superconducting quantum interference device) that %
applies $S(r,\theta)$ when the qubit is in $\ket{1}$. 
In trapped ions, Millican et al.~\cite{millicanEngineeringContinuousvariableEntanglement2025} use optimal control %
of modulated Jaynes-Cummings and anti-Jaynes-Cummings interactions to deterministically %
prepare two-mode squeezed vacuum with up to $\sim6$\,dB of squeezing. 
For our construction, a single controlled-squeezing (CS) gate prepares the ${m=2}$ codeword, %
while short sequences of conditional rotations extend to larger even~$m$.

\textbf{Choosing the right code:}
The bosonic codes we introduced differ by an integer number $m$ defining the number of legs and a positive number $r$ defining the squeezing strength. Beyond this, there exist many other \bosonic{} QECC options~\cite{ErrorCorrectionZoo}. Choosing the \textit{right} code depends on understanding the dominant noise bias (e.g., loss vs.\ dephasing), the implementable gate set and calibration complexity (conditional rotations vs.\ conditional squeezing) and other design factors. 
Our suggested code has a number-phase structure that offers increasing loss-robustness with larger $m$ while trading resilience to dephasing—precisely the trend quantified in our numeric analysis (Fig.~\ref{fig:numerical_results1}). These trade-offs should be optimized for a given platform under its measured loss/dephasing rates.

\textbf{Summary:}
We introduced a new family of rotation-symmetric bosonic codes built from coherent superpositions of oriented squeezed vacua. Analytically, the construction yields interleaved Fock-space support \({
    n \equiv 2k \ (\mathrm{mod}\ 2m)
}\) and code distance properties; operationally, a single conditional-squeezing (CS) gate prepares the $m{=}2$ instance, while short sequences of conditional rotations extend the scheme to general even $m$. Using a Knill--Laflamme violation measure, we mapped the trade-off across values of $m$ and benchmarked against cat codes, observing increased loss-robustness with larger $m$.
Finally, we analyzed potential implementations of the squeezed-vacuum codes on existing experimental platforms.
These results position the family of squeezed-vacuum codes as a practical number-phase encoding that complement existing cat and binomial families.

    
    \printbibliography

    \clearpage
    \onecolumn
    \balance
    \appendix
\section*{Supplementary Material}
\addcontentsline{toc}{section}{Supplementary Material}

\section{Full Analytical Derivation of the Conditional-Squeezing Code}
\label{sec:supp:code-derivation}

\def\normK{\Tilde{N}_k}
\def\kNonNeg{k^{*}}

Here we derive the closed-form photon-number amplitudes for the logical basis states Eq.~(\ref{eq:psi_k}).  Start from the squeezed vacuum written in the Fock basis \cite{weedbrookGaussianQuantumInformation2012} along axis angle $\theta$ (using $S(r,\theta)\equiv S(r e^{i\phi(\theta)})$ with $\phi(\theta)=2\theta+\pi$):
\begin{equation}
\label{eq:squeezed_vacuum_fock_rep}
  S(r,\theta)\ket 0
  =
    \frac1{\sqrt{\cosh r}}
    \sum_{n=0}^{\infty} 
    \frac{
        \sqrt{(2n)!}
    }{
        2^{n}n!
    }
    e^{
        in 2\theta 
    }\,
    \tanh^{n} r
    \ket{2n}.
\end{equation}
Rotating the squeezing axis by $\theta_j=\pi j/m$ introduces a phase $e^{i2n\theta_j}$ in the $n$-th term. Substituting into Eq.~(\ref{eq:psi_k}) and using the roots-of-unity filter yields
\begin{align}
\label{eq:fock_amp_derivation}
    |\psi_k\rangle
    &
    \overset{\eqref{eq:psi_k}}{=}
    \normK
    \sum_{j=0}^{m-1} 
    e^{
        i \frac{2\pi j}{m} k
    } \,
    S(r, \frac{\pi j}{m}) \,
    \ket{\vac}
    \\&
    \overset{\eqref{eq:squeezed_vacuum_fock_rep}}{=}
    \frac{
        \normK
    }{
        \sqrt{\cosh r}
    }
    \sum_{n=0}^{\infty} \, 
    (\tanh(r))^{n} \,
    \frac{
        \sqrt{(2n)!}
    }{
        2^{n}n!
    }
    \underbrace{
        \sum_{j=0}^{m-1} 
        e^{
            i \frac{2 \pi j}{m}(n+k)
        }
    }_{
        m \cdot \delta_{
            n\equiv -k\,(\mathrm{mod}\,m)
        }
    }
    \ket{2n}
\end{align}

so by defining \(\kNonNeg\) to be the first non-negative value to obey %
\(\kNonNeg := (-k)\bmod m\), we get:

\begin{align}
\label{eq:fock_amp_derivation2}
    \ket{\psi_k}
    = 
    \frac{
        \normK \, m
    }{
        \sqrt{\cosh r}
    }
    \sum_{\ell=0}^{\infty}
    \frac{
        \sqrt{
            [2(\ell m+ \kNonNeg)]!
        }
    }{
        2^{\ell m+ \kNonNeg}
        (\ell m+\kNonNeg)!
    }\,
    (\tanh r)^{\ell m+\kNonNeg}\,
    \ket{2(\ell m+\kNonNeg)}.
\end{align}
where $\normK$ is a normalization factor depending on $m,k,$ and $r$. Enforcing $\braket{\psi_k|\psi_k}=1$ %
gives %
\begin{equation}
  \normK
  = 
  \frac{\sqrt{\cosh r}}{m}\,
  \Biggl[
    \sum_{\ell=0}^{\infty}
    \frac{
        [
            2(\ell m+\kNonNeg)
        ]!
    }{
        4^{\ell m+\kNonNeg}\,
        \bigl(
            (\ell m+\kNonNeg)!
        \bigr)^2
    }\,
  \tanh^{2(\ell m+\kNonNeg)} r
  \Biggr]^{-1/2}.
  \label{eq:norm-Nk}
\end{equation}

Equation~(\ref{eq:fock_amp_derivation2}) makes explicit the interleaved Fock-space support separated by $2m$ photons, i.e., %
\({
    n\equiv 2k\ \! (\mathrm{mod}\ 2m)
}\). %
This structure implies that up to $t<m$ loss events map the two logical codewords into disjoint error subspaces, enabling their (in-principle) correctability with a suitable recovery procedure.

It is also useful to provide an analytical expression for the mean photon number of the rotation-symmetric $k$ states from~\cref{eq:psi_k}:
\begin{equation}
  n_k 
  := 
  \bra{\psi_k}a^{\dagger}a\ket{\psi_k}
  = 
  \frac{
    \displaystyle\sum_{\ell=0}^{\infty} 
    2(\ell m + \kNonNeg)\,
    c_{k \ell}
  }{
    \displaystyle\sum_{\ell=0}^{\infty} 
    c_{k \ell}
  },
  \quad %
  c_{k \ell} 
  := 
  \frac{
    [2(\ell m+\kNonNeg)]!
  }{
    4^{\ell m+\kNonNeg}\,
    \bigl(
      (\ell m+\kNonNeg)!
    \bigr)^2
  }\,
  (\tanh r)^{2(\ell m+\kNonNeg)}
  \label{eq:nk-mean}
\end{equation}

\section{Knill--Laflamme Noise Analysis for Bosonic Codes}
\label{sec:supp:kl}

For the pure-loss (attenuator) channel with Kraus operators
\(
    K_{j}^{\text{loss}}
    =
    \sqrt{
        \frac{
            \gamma^{j}
        }{
            j!
        }
    }
    \,
    (1-\gamma)^{
        \frac{n}{2}
    }\,
    a^{j}
\) %
(\cref{eq:noise_channels:loss_dephasing:krauss_rep}),
working on a Fock space truncated to \(\{|0\rangle,\dots,|n_{\max}\rangle\}\) makes the series \emph{finite}: 
\(K_{j}^{\text{loss}}\) annihilates \(|n\rangle\) whenever \(j>n\), so keeping all \(j\le n_{\max}\) yields 
\(
    \sum_{j=0}^{n_{\max}}
    {K_{j}^{\text{loss}}}^\dagger
    K_{j}^{\text{loss}}
    =
    \id
\) on the truncated space (up to numerical precision).

For number dephasing with Kraus operators
\(
    K_{j}^{{\text{dephasing}}}(\gamma)
    =
    \sqrt{
        \frac{
            \gamma^j
        }{
            j!
        }
    }\,
    e^{
        -
        \frac{\gamma}{2}
        n ^2
    }
    \,
     n^j
\),
the series is infinite even at finite cutoff.
We choose the minimal \(j_{\max}\) such that the completeness defect
\(
    \Delta
    =
    \id
    -
    \sum_{j=0}^{j_{\max}}
    {K_{j}^{\text{dephasing}}}^\dagger
    K_{j}^{\text{dephasing}}
\)
satisfies \(\|\Delta\|_\infty\le\varepsilon\) (machine precision in our runs).

\section{Ideal Number-Phase Codes}
\label{sec:supp:ideal_phase_number}

\paragraph{Rotation-symmetric codes and ideal number-phase codes.}
In the rotation-symmetric framework of \citet{Grimsmo2020}, order-$N$ bosonic codes are constructed by distributing quantum information across $N$ equally spaced phase-space directions.
The ideal number-phase codes represent the limit where the phase uncertainty vanishes completely while maintaining a fixed Fock-space grid structure.
These codes occupy discrete number states $n\equiv\ell\ (\mathrm{mod}\ N)$ for each computational basis state $\ket{\ell_N}$, achieving perfect phase localization at $N$ distinct angles.

\paragraph{Connection to squeezed-vacuum codes.}
Our squeezed-vacuum codes approach the ideal number-phase limit as the squeezing parameter $r\to\infty$.
Taking the $N=2$ case as an example, a single squeezed vacuum along angle $\theta$ has the Fock expansion of \cref{eq:squeezed_vacuum_fock_rep},
which contains only even photon numbers.
The two orthogonal primitives $\ket{\xi_{0}}\equiv S(r,0)\ket{0}$ and $\ket{\xi_{\pi/2}}\equiv S(r,\tfrac{\pi}{2})\ket{0}$ can be combined to form
\begin{equation}
\label{eq:our-N2-code}
\ket{0_{N=2}}\ \propto\ \ket{\xi_{0}}+\ket{\xi_{\pi/2}},
\qquad
\ket{1_{N=2}}\ \propto\ \ket{\xi_{0}}-\ket{\xi_{\pi/2}},
\end{equation}
with Fock-space expansions:
\begin{equation}
\label{eq:our-N2-code-Fock}
    \ket{\xi_{0}} \pm \ket{\xi_{\pi/2}}
\ \propto\
\sum_{n=0}^{\infty}\!\frac{\sqrt{(2n)!}}{2^n n!}\,\tanh^n r\ \big(1\pm e^{i n\pi}\big)\ket{2n},
\end{equation}
where the ``$+$'' combination selects $n\equiv0\ (\mathrm{mod}\ 4)$ and the ``$-$'' combination selects $n\equiv2\ (\mathrm{mod}\ 4)$.
This realizes the characteristic ``jumps of $4$'' Fock-space structure of an order-$2$ rotation code; %
but with only two rotated primitives instead of the four that other codes typically require %
(For example, the $4$-legged cat codes require four displaced coherent states in order to be considered an order-$2$ rotation code).

As $r\to\infty$, the two squeezed-vacuum primitives in the 2-legged code become increasingly localized in phase space around their respective angles $\theta=0$ and $\theta=\pi/2$ (noting that each primitive has two diametrically opposed lobes).
In this limit, the dual-basis states---the primitives themselves---sharpen into perfect phase eigenstates at these two angles.
Simultaneously, the Fock-grid spacing remains fixed at intervals of $4$ photons.
Thus, as $r\to\infty$, our 2-legged construction \eqref{eq:our-N2-code} approaches the ideal order-$N{=}2$ number-phase code defined in \cite{Grimsmo2020}: states with vanishing phase uncertainty and discrete support on $n\equiv\ell\ (\mathrm{mod}\ 4)$ in the Fock basis.

\paragraph{General $m$-legged codes.}
This relation extends to arbitrary $m$-legged squeezed-vacuum codes.
For an $m$-legged code, we rotate the squeezed-vacuum primitive $S(r,\phi)\ket{0}$ in steps of $\pi/m$ (\cref{eq:family_def}) to generate $m$ distinct primitives.
Since the squeezed vacuum has even Fock parity (only even photon numbers), the code words acquire support on $n=2km$ and $n=(2k+1)m$ for the logical states $\ket{0}$ and $\ket{1}$, respectively.
This realizes an order-$N{=}m$ rotation-symmetric code with Fock spacing of $2m$.
As $r\to\infty$, the $m$ primitives sharpen into phase eigenstates at $m$ equally spaced angles, thereby approaching the ideal order-$N{=}m$ number-phase code structure.

\medskip
\noindent\textit{Remark on the number of primitives.}
The rotation-code projector formalism \cite{Grimsmo2020} formally requires $2N$ rotated copies of a generic primitive to construct an order-$N$ code.
However, for even-parity primitives like the squeezed vacuum, rotations by $\pi$ are redundant---they return the same state up to a global phase.
Consequently, only $N$ distinct angles are physically needed when using rotated squeezed-vacuum primitives.

\end{document}